\documentclass[11pt,letter]{article}
\usepackage[utf8]{inputenc}
\usepackage[margin=1in]{geometry}
\usepackage{amsmath,amssymb, lmodern, xcolor, graphicx, booktabs, caption, cite}
\usepackage[
    colorlinks=true,
    linkcolor=teal,      % Color of internal links
    citecolor=red,       % Color of citation links
    urlcolor=blue,    % Color of external URLs
    linktoc=page         % Links ONLY the page numbers in the Table of Contents
]{hyperref}
\let\vec\mathbf

\newcommand{\es}[2] {\begin{equation} \label{#1} \begin{split} #2 \end{split} \end{equation}}
\newcommand{\D}{{\rm d}}

\newcommand{\bphi}{\bar{\phi}}

\title{\vspace{-1.5cm} \Large \bfseries Inflationary Origins of Not-Quite-Primordial Black Holes}
\author{Soubhik Kumar\footnote{soubhik.kumar@tufts.edu}~ and Gabriella Toryk\footnote{gabriella.toryk@tufts.edu}}

\date{}

\begin{document}

\maketitle
\vspace{-1.5em}
{\small {\it \noindent Institute of Cosmology, Department of Physics and Astronomy, Tufts University, Medford, MA 02155, USA}}

\vspace{2.5em}

% Abstract section with horizontal dividers
\noindent\rule{\textwidth}{0.6pt}
\vspace{0.5em}

\noindent{\bfseries Abstract}\\[0.5em]
Primordial Black Holes (PBHs) provide compelling motivation to study an enhanced primordial power spectrum ($\Delta_{\mathcal{R}}^2$) on scales much smaller than a megaparsec.
In single-field inflation models, these enhancements can be generated from extremely flat regions in the potential, often requiring substantial tuning of parameters. 
The recently proposed Not-Quite-Primordial Black Hole (NQPBH) scenario, motivated by observations of high redshift black holes by the James Webb Space Telescope, instead considers more moderate enhancements of the primordial scalar power spectrum, $\Delta_{\mathcal{R}}^2\sim10^{-7}$.
By focusing on two inflationary mechanisms, we show that these moderate enhancements can be generated with significantly reduced parameter fine tuning. 
First, we consider a Gaussian feature superimposed on an otherwise smooth background potential, while also providing a microscopic realization of the feature from a two-field model.
Second, we investigate a heavy particle-production mechanism in which non-adiabatic production of an additional scalar field sources enhanced $\Delta_{\mathcal{R}}^2$. 
For the first scenario, we find ${\cal O}(10\%)$ tuning of parameters, while for the second essentially no fine tuning is required.
Both scenarios can explain the observed abundance of `Little Red Dots' for plausible choices of astrophysical parameters.
More generally, these mechanisms provide a way to generate supermassive black hole seeds in the early universe, with possible imprints on dark matter substructure.

\noindent\rule{\textwidth}{0.6pt}
\newpage
\tableofcontents
\section{Introduction}
Primordial Black Holes (PBHs) have long been studied as possible consequences of enhanced density perturbations in the early universe~\cite{Zeldovich:1967lct, Hawking:1971ei, Carr:1974nx, Carr:1975qj}.
Unlike black holes formed from stellar collapse, PBHs can form from sufficiently overdense primordial fluctuations that undergo gravitational collapse after reentering the cosmological horizon. 
The potential implications of PBHs for dark matter, gravitational wave observations, and early universe cosmology have motivated growing interest in studying inflationary mechanisms that could generate the overdensities needed for their formation~\cite{Sasaki:2018dmp, Carr:2020gox, Green:2020jor, Carr:2020xqk}. 
While observations of the Cosmic Microwave Background (CMB) constrain the scalar power spectrum to a size of approximately $\Delta_{\mathcal{R}}^2 \sim 10^{-9}$ on scales bigger than a megaparsec~\cite{Planck:2018jri, Bird:2010mp}, substantially larger perturbations may be generated on smaller scales without directly violating these constraints. 

In single-field inflation models, localized enhancements of $\Delta_{\mathcal{R}}^2$ can arise when the inflaton field departs from its usual slow-roll evolution, during which the inflaton velocity rapidly slows and curvature perturbations can grow on superhorizon scales~\cite{Leach:2001zf, Kinney:2005vj, Germani:2017bcs, Motohashi:2017kbs, Byrnes:2018txb}. 
Such behavior can be generated by localized features in an otherwise smooth inflationary potential~\cite{Starobinsky:1992ts, Ivanov:1994pa, Garcia-Bellido:2017mdw, Mishra:2019pzq}. 
However, producing enhancements large enough for appreciable PBH formation, typically requiring $\Delta_{\mathcal{R}}^2\sim10^{-2}$ at the relevant scales, can demand substantial fine tuning of the parameters of the inflaton potential~\cite{Hertzberg:2017dkh, Sasaki:2018dmp, Cole:2023wyx}. 
Small changes in the shape or amplitude of the feature can significantly alter the background evolution and the resulting $\Delta_{\mathcal{R}}^2$.
This, in turn, affects the abundance of PBHs exponentially.

Recently, the Not-Quite-Primordial Black Hole (NQPBH) scenario~\cite{Qin:2025ymc} has been proposed as an alternative connection between primordial fluctuations and the formation of massive black holes in the early universe.
Motivated in part by the high redshift, compact red objects (`Little Red Dots') observed by the James Webb Space Telescope (JWST)~\cite{Matthee:2023utn, Greene:2024phl, Kokorev:2024kqm, Akins:2025iqo, Kocevski:2025tft}, this scenario considers primordial enhancements that are considerably smaller than those typically required for PBH formation. 
Rather than collapsing into black holes immediately after horizon reentry, these enhanced perturbations increase the abundance of early dark matter halos.
In these halos, baryonic gas can undergo atomic cooling and massive black hole seeds may form via the `direct collapse' mechanism~\cite{Bromm:2002hb, Begelman:2006db, Lodato:2006hw}. 
In particular, enhancements of order $\Delta_{\mathcal{R}}^2\sim10^{-7}$ may lead to such direct collapse black hole seeds~\cite{Qin:2025ymc}, without requiring the much larger fluctuations associated with conventional PBH formation.

Given the more moderate fluctuations required for the NQPBH scenario, one can ask whether these moderate enhancements can be generated with less fine tuning than the enhancements needed for PBH production in single-field models.
In this work, we investigate two mechanisms for generating localized enhancements of the primordial scalar power spectrum. 
We first consider a single-field inflation model with a localized Gaussian feature~\cite{Mishra:2019pzq} superimposed on an otherwise smooth potential, studying both a `dip' and a `bump' feature. 
We also consider a realization in which an effective dip feature arises through the coupling of the inflaton to an additional scalar field. 
The resulting scalar power spectra are obtained by numerically evolving the inflationary background and solving the Mukhanov-Sasaki equation, allowing us to capture the perturbation dynamics beyond the slow-roll approximation.

As a second mechanism, we study particle production during inflation~\cite{Chung:1999ve, Romano:2008rr, Barnaby:2009mc, Pearce:2017bdc}, in which the inflaton couples to an additional scalar field whose effective mass vanishes near a particular inflaton field value. 
The resulting non-adiabatic particle production can source additional inflaton fluctuations and generate a localized enhancement of the power spectrum. 
Unlike the Gaussian dip/bump scenario, where the enhancement is driven primarily by a modification of the background inflaton dynamics, in this scenario the produced heavy particles source inflaton perturbations and enhance $\Delta_{\mathcal{R}}^2$.

We quantify how sensitively the peak value of $\Delta_{\mathcal{R}}^2$ depends on the relevant model parameters.
This can be interpreted as how much fine tuning of parameters is required to achieve a desired value of $\Delta_{\mathcal{R}}^2$.
For the dip/bump scenario, we find ${\cal O}(10\%)$ tuning is required, while for the particle production scenario the tuning is even more mild.
We also evaluate the consequences of these mechanisms for early structure formation. 
Using the matter power spectra resulting from $\Delta_{\mathcal{R}}^2$, we calculate halo mass functions and estimate the abundance of black hole seeds formed in sufficiently massive halos. 
We then consider the growth of these seeds through accretion to show that they can lead to comoving number densities $\sim 10^{-4}\,{\rm Mpc}^{-3}$ of accreting black holes with Eddington luminosities $L\gtrsim 10^{45}$~erg/s at $z\simeq 6$.
This is comparable to the observed abundance of LRDs at $z\sim 4-8$, $\sim 10^{-5}-10^{-4}\,{\rm Mpc}^{-3}$~\cite{Matthee:2023utn, Kokorev:2024kqm, Akins:2025iqo, Kocevski:2025tft, Pizzati:2024prz}, although a precise comparison depends on the LRD selection criteria and on the bolometric corrections used to infer their luminosities and black hole masses~\cite{Greene:2025njv, Ma:2025wmv}.

The rest of this work is organized as follows.
In Sec.~\ref{sec:single_field}, we focus on the Gaussian dip/bump scenario, computing $\Delta_{\mathcal{R}}^2$, the halo mass functions and the black hole abundance, and quantifying the required fine tuning.  
In Sec.~\ref{sec:pp}, we repeat these steps for the heavy particle production scenario.
We conclude in Sec.~\ref{sec:conc}.

\section{Single Field Realization}\label{sec:single_field}
It is well known that localized features on a single-field inflaton potential can enhance the primordial power spectrum~\cite{Starobinsky:1992ts, Ivanov:1994pa, Garcia-Bellido:2017mdw, Ezquiaga:2017fvi, Germani:2017bcs, Motohashi:2017kbs, Ballesteros:2017fsr, Byrnes:2018txb}.
To illustrate this, we consider the parametrization discussed in~\cite{Mishra:2019pzq},
\es{eq:pot}{
V(\phi)  = V_b(\phi) \left(1 + A \exp\left(-{(\phi-\phi_{d})^2 \over 2\sigma_d^2}\right)\right).
}
A positive (negative) $A$ leads to a bump (dip) on the potential, with $\phi_d$ denoting the location of the feature and $\sigma_d$ denoting its sharpness.
The `base' inflaton potential is denoted by $V_b(\phi)$.
While the specific form of $V_b(\phi)$ is not important for the purpose of this work, for concreteness, we choose~\cite{Kachru:2003sx, Martin:2013tda, Kallosh:2019eeu}
\es{eq:Vb}{
V_b(\phi) = V_0 {\phi^2 \over \phi^2 + M^2}.
}
This potential leads to a spectral tilt $n_s$ and tensor-to-scalar ratio $r$ consistent with the recent cosmological data; see, e.g.,~\cite{Balkenhol:2025wms}.

\subsection{Power Spectrum}
In the context of the potential in Eq.~\eqref{eq:pot}, the slow-roll approximation does not give an accurate power spectrum of the primordial scalar perturbations for the parameter choices discussed below.
Accordingly, we numerically solve the Mukhanov-Sasaki (MS) equation (see Appendix~\ref{sec:ms_review} for a review) to derive the power spectrum,
\es{eq:ms_R}{
{\cal R}'' + 2{Z' \over Z} {\cal R}'+ k^2 {\cal R} = 0.
}
or equivalently,
\es{eq:ms}{
v'' + \left(k^2 - {Z'' \over Z}\right)v = 0.
}
Here ${\cal R}$ is the gauge invariant comoving curvature perturbation; $Z = a \bar{\phi}'/{\cal H}$ is determined in terms of the scale factor $a$, the homogeneous inflaton field $\bar{\phi}$, and the conformal Hubble rate ${\cal H}$; $v=Z{\cal R}$; and primes denote derivatives with respect to conformal time $\tau$.
Thus, the inflaton potential $V(\phi)$ influences ${\cal R}$ via $Z$ and its time derivatives.
To determine $Z$, we solve the homogeneous inflaton equation of motion (EOM), simultaneously with Eq.~\eqref{eq:ms_R}.
We write the inflaton EOM as,
\es{}{
\bphi_{,NN} + (3-\epsilon)\bphi_{,N} + (3-\epsilon){\partial V/\partial \bphi \over V} = 0,
}
where we have used the $e$-folding variable $N$, $\D \tau = \D N /{\cal H}$, along with $\bphi'=\bphi_{,N}{\cal H}$ and $\bphi''={\cal H}^2\bphi_{,NN}+\bphi_{,N}{\cal H}'$.
Here $\epsilon = (\bphi_{,N})^2/2$, ${\cal H}^2 = a^2 V/(3-\epsilon)$, $8\pi G_N =1$ with $G_N$ the Newton's constant, and $X_{,N} \equiv \D X/\D N$.
Denoting $\epsilon_2 = \D \ln \epsilon/\D N = 2 \bphi_{,NN}/\bphi_{,N}$, Eq.~\eqref{eq:ms_R} can then be written more compactly as,
\es{eq:ms_compact}{
{\cal R}_{,NN}+(3-\epsilon+\epsilon_2){\cal R}_{,N} + {k^2 \over {\cal H}^2}{\cal R} = 0.
}
To solve Eq.~\eqref{eq:ms_compact} numerically, we need to specify appropriate initial conditions when cosmological modes are far inside the horizon.
We use the alternate form of  the MS equation~\eqref{eq:ms} and rewrite the parenthesis as,
\es{}{
k^2 - {Z'' \over Z} = k^2 - {\cal H}^2(2-\epsilon+3\epsilon_2/2+\epsilon_2^2/4-\epsilon\epsilon_2/2+(1/2)\D\epsilon_2/\D N) \equiv k^2 - {\cal H}^2 F(N).
}
To evaluate the time evolution for each $k$-mode, we choose an initial time when $k^2 \gg {\cal H}^2 |F(N)|$.
Then, the Bunch-Davies initial condition for $v$ can be chosen $v_{\rm ini}(k,\tau)= \exp(-i k\tau)/\sqrt{2k}$.
In terms of this, the initial conditions for ${\cal R}$ are given by,
\es{}{
{\cal R}_{\rm ini} &= {v_{\rm ini} \over Z_{\rm ini}},\\
{\cal R}_{,N\rm ini} &= {v_{\rm ini} \over {\cal H}_{\rm ini} Z_{\rm ini}}\left(-ik-{\cal H}_{\rm ini}(1+\epsilon_{2,\rm ini}/2)\right).
}
Without any loss of generality, we can set $v_{\rm ini} = 1/\sqrt{2k}$ by performing a phase rotation at the initial time, and $Z_{\rm ini} = a_{\rm ini} \bphi_{,N \rm ini}$.

\begin{figure}
    \centering
    \includegraphics[width=0.8\linewidth]{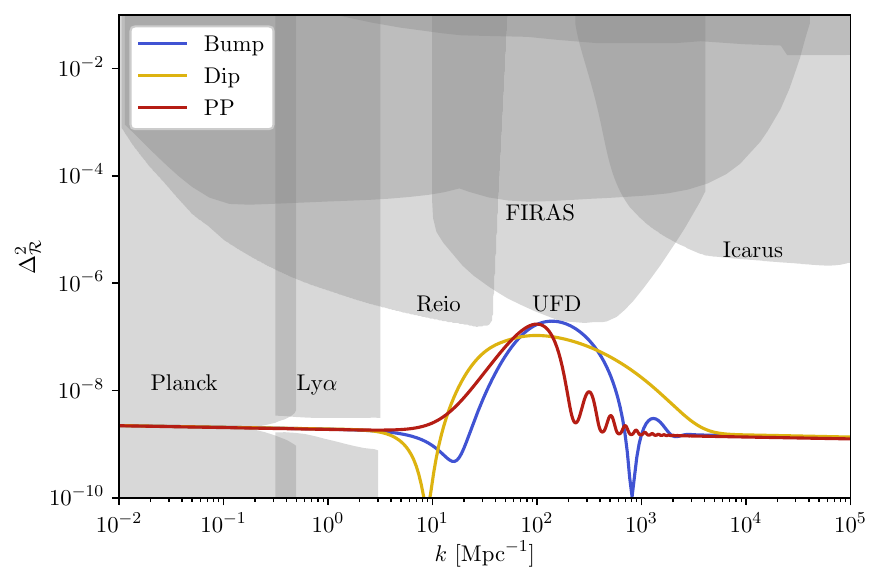}
    \caption{The primordial power spectrum from the model in Eqs.~\eqref{eq:pot} and~\eqref{eq:two_field_pp} for the fiducial parameters for the bump (blue), the dip (yellow), and particle production (red) scenarios given in the text. The particle production (PP) scenario is discussed in Sec.~\ref{sec:pp}.
    See the text for the explanations of the various constraints shown in gray. 
    }
    \label{fig:PS_bump}
\end{figure}
In Fig.~\ref{fig:PS_bump}, we show the resulting power spectrum $\Delta_{\cal R}^2(k) = k^3 \langle {\cal R}(\vec{k}) {\cal R}(-\vec{k}) \rangle/(2\pi^2)$, for both a bump and a dip on the inflaton potential.
We choose $V_0 = 7\times 10^{-11}$, $M=1/2$, $\phi_{\rm ini} = 3.22$ (in Planck units) along with the following set of parameters for the bump and dip scenarios:
\[
\begin{array}{lccc}
\toprule
 & A & \phi_d & \sigma_d \\
\midrule
\text{Bump} & 1.94\times10^{-4}  & 2.953 & 0.007 \\
\text{Dip}  & -3.78\times10^{-4} & 2.966 & 0.014 \\
\bottomrule
\end{array}
\]
Note, $\phi_d$ is chosen to ensure the peak appears at approximately $k\approx 100$/Mpc, and as such it can be varied.
We also note that for these two choices, inflation ends at different times, with a total $e$-folding of $50.7$ for the dip and $49.6$ for the bump, after a pivot scale $0.05$/Mpc exits the horizon.
A larger $\sigma_d$ for the dip, compared to the bump, leads to a broader enhancement of the power spectrum in Fig.~\ref{fig:PS_bump}.
These parameters are chosen to be consistent with the existing constraints from CMB anisotropies (Planck)~\cite{Planck:2018jri}, Lyman-$\alpha$ (Ly$\alpha$)~\cite{Bird:2010mp}, CMB spectral distortion (FIRAS)~\cite{Fixsen:1996nj, Chluba:2012we}, not having too early reionization (Reio)~\cite{Qin:2025ymc}, heating of ultra-faint dwarf galaxies due to dark matter clumps (UFD)~\cite{Graham:2024hah}, and microlensing of the Icarus star~\cite{Bringmann:2025cht}. 
We note that the last three (Reio, UFD, and Icarus) are subject to late-time astrophysical uncertainties, while the FIRAS bound is comparatively more robust, although weaker in the regime of interest.
Furthermore, these three astrophysical bounds are derived for different shapes of $\Delta_{\cal R}^2$ enhancements compared to the ones shown in Fig.~\ref{fig:PS_bump}.
Additional constraints for similar wavenumbers are discussed in~\cite{Sabti:2021unj, Dekker:2024nkb, Ando:2024ghr,  Munoz:2026diy}.
Since our goal in the present work is to illustrate inflationary mechanisms that can seed supermassive black holes, we leave a detailed re-interpretation of all these bounds for the specific $\Delta_{\cal R}^2$ enhancement for a future work.

\subsubsection{A Localized Dip from a Heavy Field}
\begin{figure}
    \centering
    \includegraphics[width=0.45\linewidth]{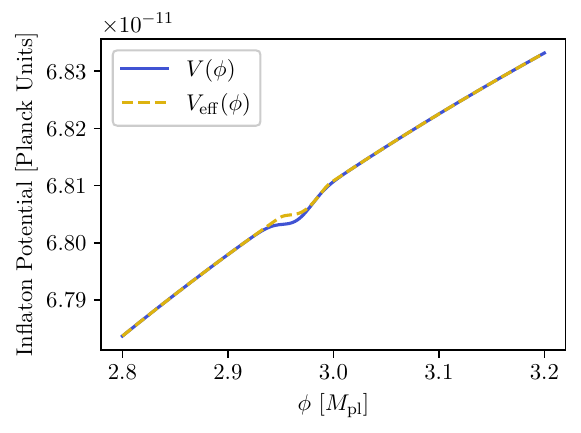}
    \includegraphics[width=0.45\linewidth]{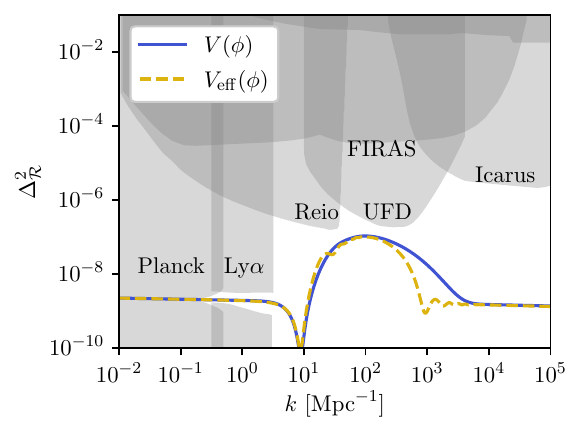}
    \caption{(Left) Comparison of inflaton potential in eq.~\eqref{eq:pot} with that in eqs.~\eqref{eq:pot_2} and~\eqref{eq:pot_3}. We choose the various parameters as: $\lambda_\chi=0.01$, $g=6.55\times 10^{-3}$, $v=1.63\times 10^{-3}$, $\phi_c=2.973$, along with the parameters for the dip given in the table above. (Right) Comparison of the resulting power spectrum.
    }
    \label{fig:inf_pot}
\end{figure}

By itself, the feature in Eq.~\eqref{eq:pot} is ad-hoc.
However, a dip ($A<0$) can arise from a more realistic model involving a second, heavier field $\chi$.
Suppose the two field potential is given by,
\es{eq:two_field}{
{\cal V}(\phi, \chi) = V_b(\phi) + {1 \over 2} g^2 (\phi - \phi_c)^2 \chi^2 + {\lambda_\chi \over 4} (\chi^2 - v^2)^2 - {\lambda_\chi \over 4} v^4.
}
Without the inflaton coupling, the mass of $\chi$ is of the order $\sqrt{\lambda_\chi}v$, which we assume to be much larger than the inflationary Hubble scale $H$.
However, the $\phi-\chi$ coupling determines where the heavy field gets localized and the associated vacuum energy.
In particular, for $\phi > \phi_c+\sqrt{\lambda_\chi} v/g$, $\chi$ is localized at $\chi=0$.
On the other hand, as the inflaton crosses the threshold $\phi=\phi_c+\sqrt{\lambda_\chi} v/g$, the $Z_2$ symmetry in $\chi$ spontaneously breaks and $\chi$ gets localized at a different minimum satisfying $\chi\neq 0$.
Eventually, $\phi$ crosses the other threshold $\phi=\phi_c-\sqrt{\lambda_\chi} v/g$ and the symmetry gets restored with $\chi$ again getting localized at $\chi=0$.
This phase transition in the $\chi$ sector creates a dip on the effective single field inflaton potential $V_{\rm eff}(\phi)$.
To see this, we integrate out the $\chi$ field at tree-level for $g|\phi-\phi_c| < \sqrt{\lambda_\chi} v$, which leads to
\es{eq:pot_2}{
V_{\rm eff}(\phi) = -{g^4 (\phi - \phi_c)^4 \over 4 \lambda_\chi} + {1 \over 2} g^2 (\phi - \phi_c)^2 v^2 - {\lambda_\chi \over 4} v^4+ V_b(\phi)~{\rm for}~g|\phi-\phi_c| < \sqrt{\lambda_\chi} v,
}
while for $g|\phi-\phi_c| > \sqrt{\lambda_\chi} v$, integrating out $\chi$ leads to
\es{eq:pot_3}{
V_{\rm eff}(\phi) = V_b(\phi)~{\rm for}~g|\phi-\phi_c| > \sqrt{\lambda_\chi} v.
}
In the left panel of Fig.~\ref{fig:inf_pot}, we compare the potential in Eq.~\eqref{eq:pot} with that in Eqs.~\eqref{eq:pot_2} and~\eqref{eq:pot_3}, for the specified $g, \lambda_\chi, v$.
The right panel of Fig.~\ref{fig:inf_pot} shows that $V_{\rm eff}$ qualitatively reproduces the height and the overall shape of the power spectrum.

We note that at the boundaries $g|\phi-\phi_c| = \sqrt{\lambda_\chi} v$, the $\chi$ mass passes through zero.
Thus the evolution becomes non-adiabatic for a brief amount of time, $\Delta t \sim 1/\mu_\chi = 1/(2\sqrt{\lambda_\chi} v g \dot{\phi}_0)^{1/3}$.
For our chosen parameters $\mu_\chi \sim 10 H$ and this time scale is much smaller than the Hubble time.
Furthermore, the energy density of the produced $\chi$ particles, $\mu_\chi^4$, is much smaller than the inflaton kinetic energy.
Thus we can ignore the backreaction of these particles on the inflaton dynamics.
However, given the $Z_2$ symmetry, there will be transient domain wall formation as well.
A small bias term $\propto \chi$ in the potential can remove these, as well as the small non-adiabatic production discussed above.

\subsection{Halo Mass Function and Black Hole Abundance}\label{sec:mass_func}
\begin{figure}
    \centering
    \includegraphics[width=0.6\linewidth]{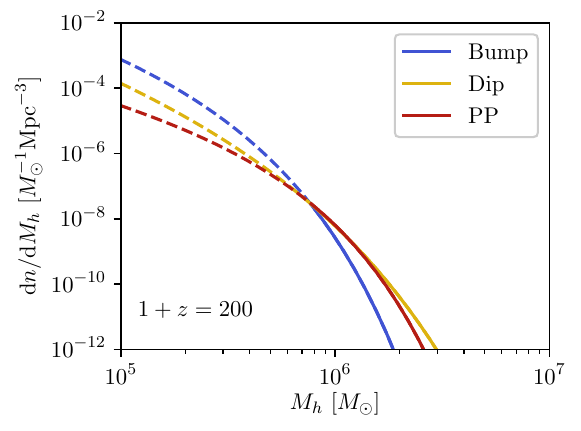}
    \caption{The halo mass function at $1+z=200$ for the bump and the dip scenarios given in Fig.~\ref{fig:PS_bump}. In $\Lambda$CDM, halos of these masses form much later, so their abundance at this redshift is exponentially smaller and do not appear on this plot. We also show the HMF for the particle production (PP) scenario discussed in Sec.~\ref{sec:pp}. We consider seed black hole formation only in halos with $M_h > M_{\rm cut} \approx 7.7\times 10^5 M_\odot$, corresponding to the Jeans mass of the collapsing gas; this part of the HMF is illustrated by solid lines. See the text for further details.
    }
    \label{fig:mass_func_bump_dip}
\end{figure}

We now discuss how an enhanced $\Delta_{\cal R}^2$ affects the halo mass function (HMF).
For a Gaussian field, the HMF can be computed using the standard methods.
We use a top-hat filter in position space whose Fourier transform is given by,
\es{}{
W(k; R) = {3 \over (k R)^3}\left(\sin(k R) - k R \cos(k R)\right),
}
and use it to compute the variance $\sigma^2(M_h, z)$ of the smoothed density field at redshift $z$:
\es{}{
\sigma^2(M_h, z) = \int \D \ln k~ \Delta_m^2(k, z) |W(k; R)|^2.
}
Here $\Delta_m^2(k, z)$ is the matter power spectrum at redshift $z$ computed from $\Delta_{\cal R}^2$ shown in Fig.~\ref{fig:PS_bump}.
We use the {\tt CLASS}~\cite{Lesgourgues:2011re, Blas:2011rf} transfer function with the Eisenstein-Hu transfer function~\cite{Eisenstein:1997ik} for $k > 5\times 10^3$/Mpc.
The smoothing scale $R$ and the halo mass $M_h$ are related in terms of the comoving mean matter density $R = (3 M_h / (4\pi\rho_m))^{1/3}$ with $\rho_m=3.97\times 10^{10} (\Omega_m/0.315) (h/0.674)^2 M_\odot/{\rm Mpc}^3$.
Using this variance, the HMF is given by,
\es{}{
{\D n \over \D \ln M_h} = - f_{\rm SMT}(\nu){\rho_m \over M_h}{\D \ln \sigma(M_h,z) \over \D \ln M_h}.
}
Here we have used the Sheth-Mo-Tormen mass function,
\es{}{
f_{\rm SMT}(\nu) = \bar{A} \sqrt{2a \over \pi} \nu (1 + (a  \nu^2)^{-p}) \exp(-a\nu^2 / 2)
}
with the parameters $\bar{A}=0.3222$, $p=0.3$, and $a=0.707$ and the peak height $\nu = 1.686/\sigma(M_h,z)$.
We show the resulting HMF in Fig.~\ref{fig:mass_func_bump_dip}.

These early halos can form at redshifts $1+z\sim 200$ and may provide massive seeds for supermassive black holes~\cite{Qin:2025ymc}.
We adopt the spherical collapse relation between halo mass $M_h$ and virial temperature $T_{\rm vir}$, with a virial overdensity $\Delta_{\rm vir}\simeq18\pi^2$ relative to the critical density.
Neglecting background radiation energy density, this is given by~\cite{Barkana:2000fd},
\es{}{
M_h(z, T_{\rm vir}) = 3.16\times 10^9 h^{-1}  M_\odot \left({T_{\rm vir} \over  1.98\times 10^4~{\rm K}}\right)^{3/2}  \left(0.6 \over  \mu\right)^{3/2} {1 \over \sqrt{\Omega_m (1+z)^3 + \Omega_\Lambda}},
}
where $\mu$, the mean molecular weight of gas in units of the proton mass, also depends on the temperature.
For $T > 3\times 10^4$~K, the gas is ionized and $\mu \approx 0.6$, while for $T < 10^4$~K, the gas becomes neutral and $\mu \approx 1.22$~\cite{Barkana:2000fd}.
Halos with $M_h \gtrsim 4\times 10^5 M_\odot$ therefore have $T_{\rm vir} \gtrsim 10^4$~K at $1+ z= 200$, and their gas can cool via atomic hydrogen.
As we discuss below, the cooled gas collapses only in halos whose mass exceeds the Jeans mass of the cooled gas.

After recombination and before thermal decoupling at $z\sim140$, the temperature of the unperturbed gas approximately follows the CMB temperature.
The corresponding cosmological Jeans mass is nearly independent of redshift and is given by~\cite{Barkana:2000fd},
\es{}{
M^{\rm cos}_J = 1.3 \times 10^{5} M_\odot \left({\Omega_m h^2 \over 0.14}\right)^{-1/2}.
}
Thus, baryons can fall into the DM halos with $M_h \gg M_J^{\rm cos}$ and get heated to virial temperatures by compressions and shocks.\footnote{See~\cite{Qin:2025ymc} for a discussion of the relative velocity of baryons with respect to DM which may have a small effect on the baryonic infall.}
Unlike DM, baryons can subsequently radiate energy and contract beyond the halo virial density.
How this happens depends on the composition of the baryonic gas.
At $1+z \gtrsim 200$, CMB photons inhibit molecular hydrogen formation by photodetaching the intermediate ${\rm H}^{-}$ ion~\cite{Hirata:2006bt}.
Following  Ref.~\cite{Qin:2025ymc}, we therefore assume that molecular cooling remains ineffective in the selected halos during the phase relevant for seed BH formation.\footnote{ The redshift at which this condition is satisfied is somewhat uncertain; see the End Matter of Ref.~\cite{Qin:2025ymc} and~\cite{Hirano:2015wla, Ito:2024mla}.
In this work, we adopt $1+z=200$ as a fiducial redshift where BH seeds form, although further work is needed to precisely determine the redshift above which atomic cooling is the dominant mechanism.}
If molecular cooling remains ineffective, atomic hydrogen can cool the gas below $T_{\rm vir}$.
However, the cooling rate  declines rapidly at temperatures below $10^4$~K~\cite{Barkana:2000fd,Oh:2001ex}.
The gas thus contracts approximately isothermally at a temperature $\approx 10^4$~K.
In sufficiently low-spin systems, a large central gas concentration can then develop without extensive fragmentation.
The mass of this central object grows at a rate~\cite{Inayoshi:2019fun},
\es{}{
{\D M \over \D t} \approx {c_s^3\over G} \approx {1\over 5.4} {M_J \over t_{\rm ff}}.
}
Here the free-fall time $t_{\rm ff} = \sqrt{3\pi/(32 G \rho)}$ and the Jeans mass of the collapsing gas is
\es{}{
M_J = {\pi^3 \over 3\sqrt{4\pi}}{c_s^3 \over G^{3/2} \rho^{1/2}},
}
with $c_s = (k_B T/\mu m_p)^{1/2}$ the isothermal sound speed and $\rho$ the total matter density.
Assuming this object (potentially a BH) accretes for a time $t_{\rm acc}$, its mass becomes $\approx M_{J,b} t_{\rm acc} / t_{\rm ff}$, with $M_{J,b} = f_b M_J$ the baryonic Jeans mass, defined using $f_b\equiv \Omega_b/\Omega_m\approx 0.157$.
We take this to be the initial BH mass $M_{\rm BH, ini}$.
For gas at $\approx 10^4$K, corresponding to $\mu \approx 1.22$, and at the virial density $\rho = \Delta_{\rm vir}\rho_{\rm crit}(z)$, we get $M_J \approx 7.7 \times 10^5 M_\odot$ and $M_{J,b} \approx 1.2 \times 10^5 M_\odot$.
We thus take
\es{eq:mbh_ini}{
M_{\rm BH,ini}=\eta M_{J,b},
}
with $\eta = t_{\rm acc}/t_{\rm ff}$.
We also note that a supermassive star collapses when its mass reaches $M\sim (1.5\text{-}3.3)\times 10^5 M_\odot$~\cite{Woods:2017bhg}.
For $\eta <1$, the mass in~\eqref{eq:mbh_ini} is already below this upper limit, i.e., $M_{\rm BH, ini} < (1.5\text{-}3.3)\times 10^5 M_\odot$.
We take $\eta=0.3$ as a fiducial value.

We model the spin distribution and Eddington accretion of the seed BHs as follows.
Assuming the spin distribution of the halos is independent of their masses, the comoving number density of seed black holes is 
\es{}{
n_{\rm BH} = {\cal N}_\lambda\int_{M_{\rm cut}}^\infty {\D M_h}~ {\D n \over \D M_h},
}
where ${\cal N}_{\lambda}$ characterizes the fraction of low-spin halos where the monolithic collapse would occur.
We take ${\cal N}_\lambda=0.1$, the fraction of halos with dimensionless spin parameter smaller than $\lambda_{\rm max}=0.02$~\cite{Qin:2025ymc}.
Here we only include halos with $M_h > M_{\rm cut} \equiv M_J \approx 8\times 10^5 M_\odot$, for which the cooled gas is subject to Jeans instability.
Such halos follow the mass function shown via solid lines in Fig.~\ref{fig:mass_func_bump_dip}.
For these BHs to potentially explain the LRDs, they must reach $L_{\rm Edd} > 10^{45}$~erg/s, i.e., $M_{\rm BH} > 8.0\times10^{6}M_\odot$ at $z=6$.

We use the Eddington growth factor ${\cal G}$ to determine how the BH mass grows:
\es{}{
M_{\rm BH}(z) =  {\cal G}(z) M_{\rm BH, ini} = M_{\rm BH, ini} \exp\left({f_{\rm duty} \Delta t (1-\epsilon_r)\over \epsilon_r \tau_{\rm Edd}}\right),
}
where the Eddington time ${\rm \tau}_{\rm Edd}=\sigma_T c/(4\pi G_N m_p) \approx 450$~Myr is defined in terms of the Thomson cross section $\sigma_T$, $\epsilon_r$ is the efficiency of accreted mass converting into radiation, and $f_{\rm duty}$ is the duty cycle that captures the fraction of time the BH is actively accreting.
We take $\epsilon_r = 0.1$ and $f_{\rm duty}=0.3$.
Finally, $\Delta t$ is the cosmic time interval between the initial collapse redshift $z_{\mathrm{coll}}$, which we take to be 199, and the redshift \(z\) at which the black-hole mass is evaluated:
\es{}{
\Delta t = \int_{1/200}^{1/(1+z)} {\D a \over a H_0 \sqrt{\Omega_m a^{-3} + \Omega_\Lambda}}.
}
We find that for $\eta>0.26$, the BHs indeed have Eddington luminosity $L_{\rm Edd} > 10^{45}$~erg/s at $z=6$ and hence in the range of the LRD luminosity observed by JWST~\cite{Matthee:2023utn, Greene:2024phl, Kokorev:2024kqm, Akins:2025iqo, Kocevski:2025tft}, although see~\cite{Greene:2025njv}.

Given the HMF in Fig.~\ref{fig:mass_func_bump_dip} and the parameter choices given above, we find $n_{\rm BH} \approx 2.9 \times 10^{-4}~{\rm Mpc}^{-3}$ and $4.6 \times 10^{-4}~{\rm Mpc}^{-3}$, for the bump and the dip scenarios, respectively.
For $f_{\rm duty}=0.3$, the number density of black holes that are accreting at any given time is roughly $0.3 n_{\rm BH}$.
While these numbers are close to the observed LRD number densities~\cite{Matthee:2023utn, Greene:2024phl, Kokorev:2024kqm, Akins:2025iqo, Kocevski:2025tft}, we emphasize our predictions of $n_{\rm BH}$ are sensitive to several parameters, including $M_{\rm cut}$, $N_\lambda$, and the formation redshift, and can vary by more than an order of magnitude.
A more detailed treatment of astrophysical effects is needed to precisely determine $n_{\rm BH}$.
We also note that recent empirical estimates of LRD bolometric luminosities are about an order of magnitude lower than earlier ones, implying correspondingly lower black hole masses~\cite{Greene:2025njv}, and that the LRD luminosity function appears to cut off sharply near $L_{\rm bol}\sim 10^{45}$~erg/s~\cite{Ma:2025wmv}.

\subsection{Quantifying the Fine Tuning}
The enhanced power spectrum for certain parameter choices comes from the dynamics being close to the `ultra-slow roll' (USR) regime.
In the USR regime, $\partial V/\partial\phi = 0$.
In the limit $A\ll 1$, this condition applied to eq.~\eqref{eq:pot} reads as,
\es{eq:Vprimezero}{
{\partial V_b/\partial\phi \over V_b} = A \exp\left(-{(\phi-\phi_{d})^2 \over 2\sigma_d^2}\right){\phi-\phi_d \over \sigma_d^2}.
}
The magnitude of the Gaussian contribution to the potential slope is maximized at \(|\phi-\phi_d|=\sigma_d\). For the Gaussian bump (dip), the cancellation occurs approximately at \(\phi=\phi_d+\sigma_d\) (\(\phi=\phi_d-\sigma_d\)). Substituting these values into Eq.~\eqref{eq:Vprimezero}, using Eq.~\eqref{eq:Vb}, and using \(\phi_d\gg\sigma_d\), we obtain
\begin{equation}
\frac{2M^2}{\phi_d(\phi_d^2+M^2)}
\simeq
\frac{|A|}{\sigma_d\sqrt{e}}.
\end{equation}
This relation gives an approximate condition for cancellation between the slopes of the bump/dip and $V_b$.
Near this condition, variations in the feature parameters can alter the residual gradient in the potential and affect the degree and the duration of the departure from slow-roll evolution. Since the enhancement of the scalar power spectrum depends on this evolution, we investigate how sensitively the peak amplitude responds to changes in the potential parameters.

\begin{figure}
    \centering
\includegraphics[width=0.45\linewidth]{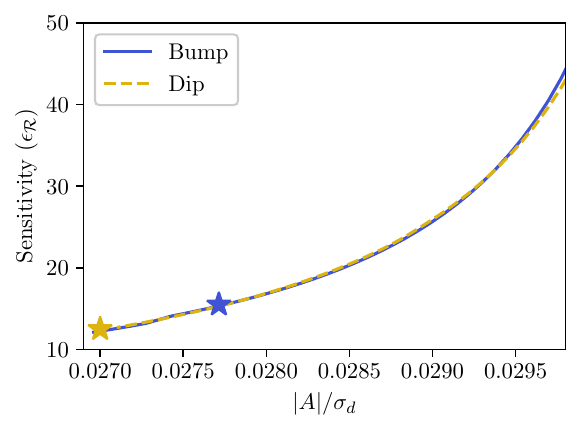}
\includegraphics[width=0.45\linewidth]{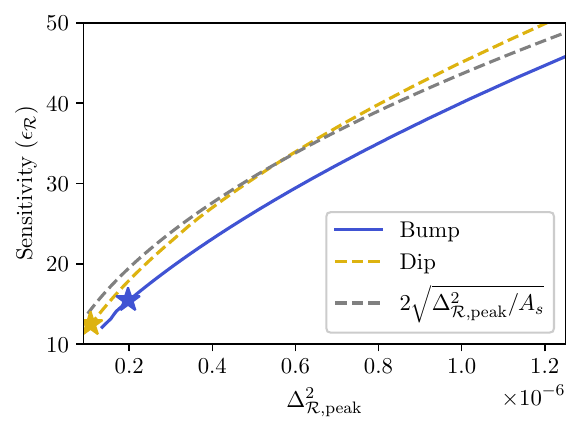}
    \caption{(Left) The sensitivity $\epsilon_{\cal R}$, defined in Eq.~\eqref{eq:sens}, of the peak of the power spectrum on the parameter $p=|A|/\sigma_d$.  (Right) $\epsilon_{\cal R}$ as a function of the peak of the power spectrum, with larger peak displaying more sensitivity. For both the panels, the stars indicate the values of $|A|/\sigma_d$ (left panel) and the peak height (right panel) for the spectra shown in Fig.~\ref{fig:PS_bump}.
    }
    \label{fig:sens}
\end{figure}

A useful measure of fine tuning is defined as~\cite{Cole:2023wyx}:
\begin{equation}
    \epsilon_{\mathcal{R},p}=\bigg\lvert\frac{\partial \ln \Delta^2_{{\cal R},\rm{peak}}}{\partial \ln |p|}\bigg\rvert.\label{eq:sens}
\end{equation}
This quantity measures the fractional change in the peak amplitude resulting from a fractional change in the parameter \(p\). A larger magnitude of \(\epsilon_{\mathcal{R},p}\) therefore corresponds to a greater sensitivity of the resulting enhancement to variations in that parameter.  Numerically we estimate this value by using finite variations about the fiducial parameter value, $p_0$. For a symmetric fractional variation \(\delta\), the logarithmic derivative can be approximated as
\begin{equation}
    \epsilon_{\mathcal{R},p}\simeq\frac{\ln(\Delta_{\cal{R},{\rm peak}}^2(p_0[1+\delta]))-\ln(\Delta_{\cal{R},{\rm peak}}^2(p_0[1-\delta]))}{\ln (p_0[1+\delta])-\ln (p_0[1-\delta])}.
\end{equation}
In the numerical analysis, we focus on variations of the ratio \(|A|/\sigma_d\), which controls the relative strength of the Gaussian feature.
We compute the peak amplitude of the scalar power spectrum across a range of values of $|A|$ and estimate the logarithmic sensitivity, while holding the remaining potential parameters fixed.
The results are shown in Fig.~\ref{fig:sens}.
We note that for the bump and dip parameter values considered in this work, $\epsilon_{\cal R}$ are approximately 16 and 12, respectively, indicating roughly $1/\epsilon_{\cal R} \approx 5{\text-}10\%$ `fine tuning'.
This is also expected based on the relation $\epsilon_{\cal R} \sim 2\sqrt{\Delta_{\cal R, {\rm peak}}^2 / A_s}$~\cite{Cole:2023wyx} (shown in gray dashed in Fig.~\ref{fig:sens}), which leads to $\epsilon_{\cal R} \approx 14-19$.
Notably, for canonical primordial black hole formation in single-field inflation models, one needs $\Delta_{\cal R, {\rm peak}}^2 \sim 10^{-2}$, leading to $\epsilon_{\cal R} \gtrsim 4\times 10^3$, indicating significantly more fine tuning.

\section{Heavy Particle Production}\label{sec:pp}
We now consider the model~\eqref{eq:two_field} with $\lambda_\chi=0$, i.e.,
\es{eq:two_field_pp}{
{\cal V}(\phi, \chi) = V_b(\phi) + {1 \over 2} g^2 (\phi - \phi_c)^2 \chi^2.
}
This model has been widely considered in the literature~\cite{Chung:1999ve, Kofman:2004yc, Romano:2008rr, Barnaby:2009mc, Green:2009ds}.
In this case, the $\chi$ field becomes massless as the inflaton crosses the critical field value $\phi_c$.
As a result, it cannot be integrated out and in fact, copious $\chi$ production takes place.
The produced $\chi$ particles backreact on the inflaton field and modify the primordial power spectrum $\Delta_{\cal R}^2$.

This contribution to the power spectrum has been computed in several works~\cite{Chung:1999ve, Romano:2008rr, Barnaby:2009mc, Barnaby:2009dd, Barnaby:2010ke}; we adopt the analytical expressions from~\cite{Pearce:2017bdc}.
The result at one-loop is given by,
\es{eq:ps_pp}{
\Delta_{\cal R}^2(k) \approx A_s \left({k\over k_p}\right)^{n_s-1}\left(1 + {({2 + \sqrt{2}) g^2}(g|\dot{\phi}_0|)^{3/2} \over 8\pi^3 H^3}f\left({k\over a_* H}\right)\right),
}
where $f(x) = (\sin(x) - {\rm SinIntegral}(x))^2 / x^3$.
We take the pivot scale to be $k_p=0.05$/Mpc, along with $A_s=2.1\times 10^{-9}$, $n_s=0.965$~\cite{Planck:2018vyg}, and $\dot{\phi}_0/H^2 \approx 62^2$ (at the particle production scale).
The function $f$ has a peak at $x=k/(a_* H)=3.346$ and so we define $a_* H = k_{\rm peak} / 3.346$.
It was also shown in~\cite{Pearce:2017bdc} that the two-loop contribution is smaller than the one-loop result above for $g^{5/2}|\dot{\phi}_0|^{1/2}/H \ll (2\pi)^3$.
For our choice of $g=0.7$, this condition is obeyed, and we show the resulting power spectrum, after also choosing $k_{\rm peak}=100$/Mpc, in Fig.~\ref{fig:PS_bump} in red (labeled `PP').

The computation of the halo mass function and the black hole abundance, given the enhanced power spectrum in Fig.~\ref{fig:PS_bump}, follows from the discussion in Section~\ref{sec:mass_func}.
We show the results in Fig.~\ref{fig:mass_func_bump_dip} in red (labeled `PP').
Repeating the computation for the comoving black hole number density, using the same assumptions as in Sec~\ref{sec:mass_func}, we find $n_{\rm BH}\approx 4.7 \times 10^{-4}~{\rm Mpc}^{-3}$.
Note, for our choice of $g$, the non-Gaussianity of density perturbations affects the tail of the HMF and can increase $n_{\rm BH}$ by a factor of several.
Given the accompanying astrophysical uncertainties in the estimate of $n_{\rm BH}$ discussed above, we do not include this non-Gaussian enhancement for simplicity.

\begin{figure}
    \centering
\includegraphics[width=0.45\linewidth]{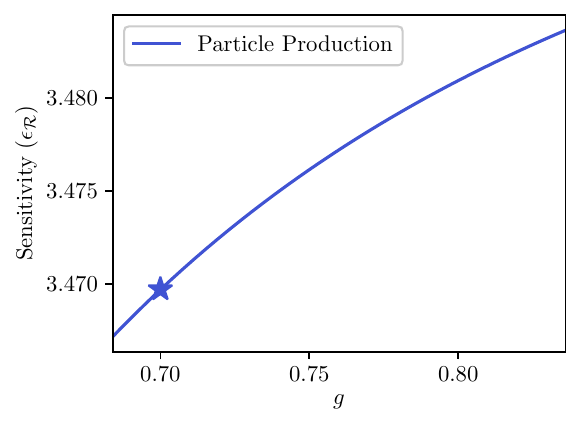}
\includegraphics[width=0.45\linewidth]{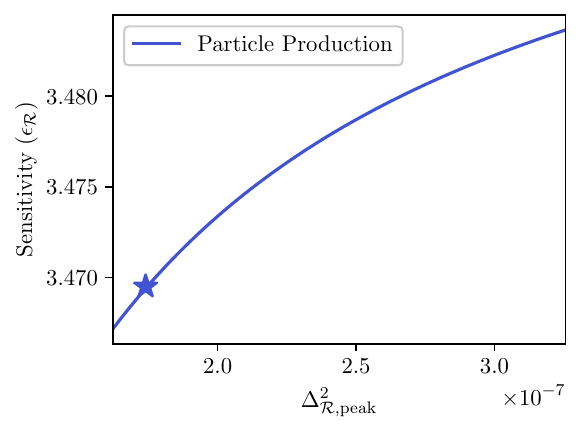}
    \caption{(Left) The sensitivity $\epsilon_{{\cal R},g}$, defined in Eq.~\eqref{eq:eps_R_g}, of the peak of the power spectrum on the parameter $g$.  (Right) $\epsilon_{\cal R}$ as a function of the peak of the power spectrum.
    Compared to Fig.~\ref{fig:sens}, the sensitivity of the peak to the model parameter is much smaller. For both the panels, the stars indicate the fiducial value $g=0.7$, which was used for the spectra in Fig.~\ref{fig:PS_bump}.
    }
    \label{fig:sens_pp}
\end{figure}

To quantify the particle-production mechanism, we consider the logarithmic sensitivity of the peak scalar power spectrum to the coupling $g$,
\begin{equation}
    \epsilon_{\mathcal{R},g}\equiv \bigg\lvert\frac{\partial\ln\Delta_{\cal{R}, {\rm peak}}^2}{\partial\ln{g}}\bigg\rvert.
    \label{eq:eps_R_g}
\end{equation}
For fixed background dynamics, the source contribution to the scalar power spectrum scales as $g^{7/2}$. Writing the total spectrum as
\begin{equation}
    \Delta_{\cal{R}}^2(k,g)=\Delta_{\cal{R},{\rm vac}}^2(k)[1+C(k)g^{7/2}],
\end{equation}
where $\Delta_{\cal{R},{\rm vac}}^2(k)$ is the standard vacuum scalar power spectrum, and $C(k)$ contains the remaining factors in the sourced contribution, including its scale dependence through the particle-production shape function. Using the one-loop expression from Eq.~\eqref{eq:ps_pp}, we can identify
\begin{equation}
    C(k)=\frac{2+\sqrt{2}}{8\pi^3}\frac{|\dot\phi_0|^{3/2}}{H^3}f\left(\frac{k}{a_* H}\right).
\end{equation}
The sensitivity follows as
\begin{equation}
    \epsilon_{\mathcal{R},g}=\frac{7}{2}\frac{C(k_{\rm peak})g^{7/2}}{1+C(k_{\rm peak})g^{7/2}}.
    \label{eq:sens_pp}
\end{equation}
In the regime where the sourced contribution dominates over the vacuum spectrum, this expression approaches $\epsilon_{\mathcal{R},g}=7/2$.
Numerically, for our fiducial coupling $g=0.7$, we obtain a peak of $\Delta^2_{\cal{R}, \rm{peak}}\simeq 1.74\times10^{-7}$ and a sensitivity of $\epsilon_{\mathcal{R},g}\simeq 3.47$, consistent with the analytical expectation.
We have further confirmed the approximately power law dependence of the peak amplitude via a scan over $0.684\leq g\leq0.836$, with a fitted logarithmic slope of $3.48$.
These results are shown in Fig.~\ref{fig:sens_pp}.

Thus, the enhancement exhibits an order unity logarithmic sensitivity to $g$, rather than a sharp dependence on the coupling.
This behavior reflects the perturbative power law scaling of the sourced contribution.
This is to be compared with the scenarios involving a bump or dip on the inflaton potential (cf. Fig.~\ref{fig:sens}), which had a much larger sensitivity on the model parameters, requiring more fine tuning, although much less compared to fine tuning associated with PBH production in single-field models.

\section{Conclusion}\label{sec:conc}
We have explored two inflationary mechanisms that enhance the power spectrum $\Delta_{\mathcal{R}}^2$ of the primordial curvature perturbation ${\cal R}$, with little to moderate fine tuning, to a level $\Delta_{\mathcal{R}}^2 \sim 10^{-7}$ at a comoving wavenumber $k \simeq 100$/Mpc.
Specifically, we found a logarithmic sensitivity $\epsilon_{\cal R} \sim 12-16$ and $\sim 3$ for the dip/bump and the particle production scenarios, respectively, with the associated fine tuning going as $1/\epsilon_{\cal R}$.
These curvature perturbations lead to enhanced dark matter overdensities which grow upon horizon reentry and collapse to form dark matter halos at redshifts $1+z \gtrsim 200$.
For suppressed molecular cooling, in low spin halos baryons undergo atomic cooling and collapse, without fragmenting, to form black holes with masses $\sim 10^{4}-10^{5} M_\odot$.
These can subsequently grow via Eddington accretion to lead to supermassive black holes in the later universe.
We have shown that the number densities of these black holes can be similar to those of the `Little Red Dots' observed by the JWST, providing a potential explanation of their origin.

There are several directions for future work.
Generally, enhanced dark matter density fluctuations can lead to early star formation as well.
The radiation from such stars, as well as accreting black holes, can potentially reionize the universe too early and modify the optical depth to reionization.
Therefore, precision cosmological data can discover or constrain such scenarios.
A dedicated analysis for the spectral shapes of $\Delta_{\cal R}^2$ considered here would be required. 
In addition, dark matter substructure resulting from these enhancements can also be probed via different methods, e.g., gravitational lensing.
It would be interesting to analyze the implications of these observables for the enhancements considered in this work.
Finally, it would be useful to perform a more detailed analysis of the initial black hole mass and its accretion history, e.g., via simulation-based methods.

\section*{Acknowledgment}
We would like to thank Wenzer Qin and Neal Weiner for useful discussions and feedback on the manuscript.
SK is supported in part by the National Science Foundation grant PHY-2609927.
This research was also supported in part by the Faculty Research Fund provided by Tufts University.
Claude was used to write part of the code used in this work, which the authors have verified.

\appendix
\section{Review of the Mukhanov-Sasaki Equation}\label{sec:ms_review}
The background equations, of which only two are independent, are given by,
\es{eq:bg}{
&\bphi''+2{\cal H}\bphi' + a^2 \frac{\partial V}{\partial \bphi} = 0,\\
&3{\cal H}^2 - 8\pi G\left(a^2 V +{1\over 2}\bphi'^2\right) = 0,\\
&{\cal H}^2 - {\cal H}' - 4\pi G \bphi'^2 = 0. 
}
Working in the Newtonian gauge, the cosmological perturbations are given by
\es{}{
 \D s^2 = a^2(\tau) \left[
-(1 + 2\Phi)\, \D \tau^2
+ (1 - 2\Psi)\,\delta_{ij}  \D x^i \D x^j
\right].
}
In the scenarios we consider the anisotropic stress is zero, leading to $\Phi=\Psi$.
The perturbations of the metric and that of the inflaton field satisfy the following equations:
\es{eq:pert}{
&3\mathcal{H}\,\Phi' + k^2\Phi 
+ 4\pi G a^2 
\left(2V\Phi
+ \frac{\partial V}{\partial \bphi}\,\delta\varphi
+\frac{\bphi'\,\delta\varphi'}{a^2}\right)= 0,\\
&\Phi'+{\cal H}\Phi-4\pi G \bphi'\delta\varphi=0,\\
&\delta\varphi''+2\mathcal{H}\,\delta\varphi' +k^2\delta\varphi+
a^2\left(2\Phi \frac{\partial V}{\partial \bphi}+\delta\varphi\,\frac{\partial^2 V}{\partial \bphi^2}  \right)
-4\bphi'\,\Phi' =0.
}
In terms of the gauge-invariant variable ${\cal R} = \Psi + {\cal H}\delta\varphi/\bphi'$, the first two of~\eqref{eq:pert}, together with~\eqref{eq:bg}, can be reduced to the well-known Mukhanov--Sasaki (MS) equation,
\es{}{
{\cal R}'' + 2{Z' \over Z} {\cal R}'+ k^2 {\cal R} = 0,
}
where $Z = a \bphi'/{\cal H}$.
To get a more conventional form, in terms of the variable $v=Z{\cal R}$, this can be rewritten as~\eqref{eq:ms}.

\paragraph{Slow-roll Approximation}
To the leading order in slow-roll approximation, $Z''/Z \approx 2/\tau^2$.
Then the MS equation admits an analytical solution.
Upon quantization, the mode function multiplying the annihilation operator is given by,
\es{}{
v_{\vec{k}}(\tau) = {1 \over \sqrt{2 k}}\left(1- {i\over k\tau}\right)e^{-ik\tau}.
}
The power spectrum of $v$ in the late time $\tau\rightarrow 0$ limit is given by, $P_v = 1 / (2k^3\tau^2)$.
Accordingly, the power spectrum of ${\cal R}$ is given by,
\es{}{
P_{\cal R} = {1 \over 2k^3\tau^2}{1 \over Z^2} = {H^4 \over 2 k^3 \dot{\bphi}^2},
}
where in the last step $H$ and $\dot{\bphi}$ are evaluated at horizon crossing $k=aH$.

\bibliographystyle{utphys}
\bibliography{refs}

@article{Balkenhol:2025wms,
    author = "Balkenhol, L. and others",
    title = "{Inflation at the End of 2025: Constraints on $r$ and $n_s$ Using the Latest CMB and BAO Data}",
    eprint = "2512.10613",
    archivePrefix = "arXiv",
    primaryClass = "astro-ph.CO",
    doi = "10.33232/001c.164435",
    month = "12",
    year = "2025"
}

@article{Mishra:2019pzq,
    author = "Mishra, Swagat S. and Sahni, Varun",
    title = "{Primordial Black Holes from a tiny bump/dip in the Inflaton potential}",
    eprint = "1911.00057",
    archivePrefix = "arXiv",
    primaryClass = "gr-qc",
    doi = "10.1088/1475-7516/2020/04/007",
    journal = "JCAP",
    volume = "04",
    pages = "007",
    year = "2020"
}

@article{Martin:2013tda,
    author = "Martin, Jerome and Ringeval, Christophe and Vennin, Vincent",
    title = "{Encyclop{\ae}dia Inflationaris}: {Opiparous Edition}",
    eprint = "1303.3787",
    archivePrefix = "arXiv",
    primaryClass = "astro-ph.CO",
    doi = "10.1016/j.dark.2024.101653",
    journal = "Phys. Dark Univ.",
    volume = "5-6",
    pages = "75--235",
    year = "2014"
}

@article{Kachru:2003sx,
    author = "Kachru, Shamit and Kallosh, Renata and Linde, Andrei D. and Maldacena, Juan Martin and McAllister, Liam P. and Trivedi, Sandip P.",
    title = "{Towards inflation in string theory}",
    eprint = "hep-th/0308055",
    archivePrefix = "arXiv",
    reportNumber = "SLAC-PUB-9669, SU-ITP-03-18, TIFR-TH-03-06",
    doi = "10.1088/1475-7516/2003/10/013",
    journal = "JCAP",
    volume = "10",
    pages = "013",
    year = "2003"
}

@article{Kallosh:2019eeu,
    author = "Kallosh, Renata and Linde, Andrei",
    title = "{B-mode Targets}",
    eprint = "1906.04729",
    archivePrefix = "arXiv",
    primaryClass = "astro-ph.CO",
    doi = "10.1016/j.physletb.2019.134970",
    journal = "Phys. Lett. B",
    volume = "798",
    pages = "134970",
    year = "2019"
}

@article{Qin:2025ymc,
    author = "Qin, Wenzer and Kumar, Soubhik and Natarajan, Priyamvada and Weiner, Neal",
    title = "{Not-Quite-Primordial Black Holes}",
    eprint = "2506.13858",
    archivePrefix = "arXiv",
    primaryClass = "astro-ph.CO",
    doi = "10.1103/zfjm-8fnt",
    journal = "Phys. Rev. Lett.",
    volume = "137",
    number = "11",
    pages = "111004",
    year = "2026"
}

@article{Bringmann:2025cht,
    author = "Bringmann, Torsten and Croon, Djuna and Sevillano Mu{\~n}oz, Sergio",
    title = "{Updated Constraints on the Primordial Power Spectrum at sub-Mpc Scales}",
    eprint = "2506.20704",
    archivePrefix = "arXiv",
    primaryClass = "astro-ph.CO",
    reportNumber = "IPPP/25/40",
    doi = "10.1103/2jhb-n2rk",
    journal = "Phys. Rev. Lett.",
    volume = "137",
    number = "6",
    pages = "061002",
    year = "2026"
}

@article{Graham:2024hah,
    author = "Graham, Peter W. and Ramani, Harikrishnan",
    title = "{Constraints on dark matter from dynamical heating of stars in ultrafaint dwarfs. II. Substructure and the primordial power spectrum}",
    eprint = "2404.01378",
    archivePrefix = "arXiv",
    primaryClass = "hep-ph",
    doi = "10.1103/PhysRevD.110.075012",
    journal = "Phys. Rev. D",
    volume = "110",
    number = "7",
    pages = "075012",
    year = "2024"
}

@article{Eisenstein:1997ik,
    author = "Eisenstein, Daniel J. and Hu, Wayne",
    title = "{Baryonic features in the matter transfer function}",
    eprint = "astro-ph/9709112",
    archivePrefix = "arXiv",
    reportNumber = "IASSNS-AST-97-51",
    doi = "10.1086/305424",
    journal = "Astrophys. J.",
    volume = "496",
    pages = "605",
    year = "1998"
}

@article{Hirata:2006bt,
    author = "Hirata, Christopher M. and Padmanabhan, Nikhil",
    title = "{Cosmological production of H(2) before the formation of the first galaxies}",
    eprint = "astro-ph/0606437",
    archivePrefix = "arXiv",
    doi = "10.1111/j.1365-2966.2006.10924.x",
    journal = "Mon. Not. Roy. Astron. Soc.",
    volume = "372",
    pages = "1175--1186",
    year = "2006"
}

@article{Barkana:2000fd,
    author = "Barkana, Rennan and Loeb, Abraham",
    title = "{In the beginning: The First sources of light and the reionization of the Universe}",
    eprint = "astro-ph/0010468",
    archivePrefix = "arXiv",
    doi = "10.1016/S0370-1573(01)00019-9",
    journal = "Phys. Rept.",
    volume = "349",
    pages = "125--238",
    year = "2001"
}

@article{Oh:2001ex,
    author = "Oh, S. Peng and Haiman, Zoltan",
    title = "{Second-generation objects in the universe: radiative cooling and collapse of halos with virial temperatures above 10{\textasciicircum}4 kelvin}",
    eprint = "astro-ph/0108071",
    archivePrefix = "arXiv",
    doi = "10.1086/339393",
    journal = "Astrophys. J.",
    volume = "569",
    pages = "558",
    year = "2002"
}

@article{Lodato:2006hw,
    author = "Lodato, Giuseppe and Natarajan, Priya",
    title = "{Supermassive black hole formation during the assembly of pre-galactic discs}",
    eprint = "astro-ph/0606159",
    archivePrefix = "arXiv",
    doi = "10.1111/j.1365-2966.2006.10801.x",
    journal = "Mon. Not. Roy. Astron. Soc.",
    volume = "371",
    pages = "1813--1823",
    year = "2006"
}

@article{Bromm:2002hb,
    author = "Bromm, Volker and Loeb, Abraham",
    title = "{Formation of the first supermassive black holes}",
    eprint = "astro-ph/0212400",
    archivePrefix = "arXiv",
    doi = "10.1086/377529",
    journal = "Astrophys. J.",
    volume = "596",
    pages = "34--46",
    year = "2003"
}

@article{Hirano:2015wla,
    author = "Hirano, Shingo and Zhu, Nick and Yoshida, Naoki and Spergel, David and Yorke, Harold W.",
    title = "{Early structure formation from primordial density fluctuations with a blue-tilted power spectrum}",
    eprint = "1504.05186",
    archivePrefix = "arXiv",
    primaryClass = "astro-ph.CO",
    doi = "10.1088/0004-637X/814/1/18",
    journal = "Astrophys. J.",
    volume = "814",
    number = "1",
    pages = "18",
    year = "2015"
}

@article{Ito:2024mla,
    author = "Ito, Mana and Omukai, Kazuyuki",
    title = "{First star formation in extremely early epochs}",
    eprint = "2405.10073",
    archivePrefix = "arXiv",
    primaryClass = "astro-ph.GA",
    doi = "10.1093/pasj/psae054",
    journal = "Publ. Astron. Soc. Jap.",
    volume = "76",
    number = "4",
    pages = "850-862",
    year = "2024"
}

@article{Woods:2017bhg,
    author = "Woods, T. E. and Heger, Alexander and Whalen, Daniel J. and Haemmerle, Lionel and Klessen, Ralf S.",
    title = "{On the Maximum Mass of Accreting Primordial Supermassive Stars}",
    eprint = "1703.07480",
    archivePrefix = "arXiv",
    primaryClass = "astro-ph.SR",
    doi = "10.3847/2041-8213/aa7412",
    journal = "Astrophys. J. Lett.",
    volume = "842",
    number = "1",
    pages = "L6",
    year = "2017"
}

@article{Matthee:2023utn,
    author = "Matthee, Jorryt and others",
    title = "{Little Red Dots: An Abundant Population of Faint Active Galactic Nuclei at z {\ensuremath{\sim}} 5 Revealed by the EIGER and FRESCO JWST Surveys}",
    eprint = "2306.05448",
    archivePrefix = "arXiv",
    primaryClass = "astro-ph.GA",
    doi = "10.3847/1538-4357/ad2345",
    journal = "Astrophys. J.",
    volume = "963",
    number = "2",
    pages = "129",
    year = "2024"
}

@article{Greene:2024phl,
    author = "Greene, Jenny E. and others",
    title = "{UNCOVER Spectroscopy Confirms the Surprising Ubiquity of Active Galactic Nuclei in Red Sources at z {\ensuremath{>}} 5}",
    doi = "10.3847/1538-4357/ad1e5f",
    journal = "Astrophys. J.",
    volume = "964",
    number = "1",
    pages = "39",
    year = "2024"
}

@article{Kokorev:2024kqm,
    author = "Kokorev, Vasily and others",
    title = "{A Census of Photometrically Selected Little Red Dots at 4 {\ensuremath{<}} z {\ensuremath{<}} 9 in JWST Blank Fields}",
    doi = "10.3847/1538-4357/ad4265",
    journal = "Astrophys. J.",
    volume = "968",
    number = "1",
    pages = "38",
    year = "2024"
}

@article{Akins:2025iqo,
    author = "Akins, Hollis B. and others",
    title = "{COSMOS-Web: The Overabundance and Physical Nature of {\textquotedblleft}Little Red Dots{\textquotedblright}{\textemdash}Implications for Early Galaxy and SMBH Assembly}",
    doi = "10.3847/1538-4357/ade984",
    journal = "Astrophys. J.",
    volume = "991",
    number = "1",
    pages = "37",
    year = "2025"
}

@article{Kocevski:2025tft,
    author = "Kocevski, Dale D. and others",
    title = "{The Rise of Faint, Red Active Galactic Nuclei at z {\ensuremath{>}} 4: A Sample of Little Red Dots in the JWST Extragalactic Legacy Fields}",
    eprint = "2404.03576",
    archivePrefix = "arXiv",
    primaryClass = "astro-ph.GA",
    doi = "10.3847/1538-4357/adbc7d",
    journal = "Astrophys. J.",
    volume = "986",
    number = "2",
    pages = "126",
    year = "2025"
}

@article{Greene:2025njv,
    author = "Greene, Jenny E. and others",
    title = "{What You See Is What You Get: Empirically Measured Bolometric Luminosities of Little Red Dots}",
    eprint = "2509.05434",
    archivePrefix = "arXiv",
    primaryClass = "astro-ph.GA",
    doi = "10.3847/1538-4357/ae1836",
    journal = "Astrophys. J.",
    volume = "996",
    number = "2",
    pages = "129",
    year = "2026"
}

@article{Cole:2023wyx,
    author = "Cole, Philippa S. and Gow, Andrew D. and Byrnes, Christian T. and Patil, Subodh P.",
    title = "{Primordial black holes from single-field inflation: a fine-tuning audit}",
    eprint = "2304.01997",
    archivePrefix = "arXiv",
    primaryClass = "astro-ph.CO",
    doi = "10.1088/1475-7516/2023/08/031",
    journal = "JCAP",
    volume = "08",
    pages = "031",
    year = "2023"
}

@article{Starobinsky:1992ts,
    author = "Starobinsky, Alexei A.",
    title = "{Spectrum of adiabatic perturbations in the universe when there are singularities in the inflation potential}",
    journal = "JETP Lett.",
    volume = "55",
    pages = "489--494",
    year = "1992"
}

@article{Ivanov:1994pa,
    author = "Ivanov, P. and Naselsky, P. and Novikov, I.",
    title = "{Inflation and primordial black holes as dark matter}",
    doi = "10.1103/PhysRevD.50.7173",
    journal = "Phys. Rev. D",
    volume = "50",
    pages = "7173--7178",
    year = "1994"
}

@article{Garcia-Bellido:2017mdw,
    author = "Garcia-Bellido, Juan and Ruiz Morales, Ester",
    title = "{Primordial black holes from single field models of inflation}",
    eprint = "1702.03901",
    archivePrefix = "arXiv",
    primaryClass = "astro-ph.CO",
    doi = "10.1016/j.dark.2017.09.007",
    journal = "Phys. Dark Univ.",
    volume = "18",
    pages = "47--54",
    year = "2017"
}

@article{Ezquiaga:2017fvi,
    author = "Ezquiaga, Jose Mar{\'\i}a and Garcia-Bellido, Juan and Ruiz Morales, Ester",
    title = "{Primordial Black Hole production in Critical Higgs Inflation}",
    eprint = "1705.04861",
    archivePrefix = "arXiv",
    primaryClass = "astro-ph.CO",
    doi = "10.1016/j.physletb.2017.11.039",
    journal = "Phys. Lett. B",
    volume = "776",
    pages = "345--349",
    year = "2018"
}

@article{Germani:2017bcs,
    author = "Germani, Cristiano and Prokopec, Tomislav",
    title = "{On primordial black holes from an inflection point}",
    eprint = "1706.04226",
    archivePrefix = "arXiv",
    primaryClass = "astro-ph.CO",
    doi = "10.1016/j.dark.2017.09.001",
    journal = "Phys. Dark Univ.",
    volume = "18",
    pages = "6--10",
    year = "2017"
}

@article{Motohashi:2017kbs,
    author = "Motohashi, Hayato and Hu, Wayne",
    title = "{Primordial Black Holes and Slow-Roll Violation}",
    eprint = "1706.06784",
    archivePrefix = "arXiv",
    primaryClass = "astro-ph.CO",
    doi = "10.1103/PhysRevD.96.063503",
    journal = "Phys. Rev. D",
    volume = "96",
    number = "6",
    pages = "063503",
    year = "2017"
}

@article{Ballesteros:2017fsr,
    author = "Ballesteros, Guillermo and Taoso, Marco",
    title = "{Primordial black hole dark matter from single field inflation}",
    eprint = "1709.05565",
    archivePrefix = "arXiv",
    primaryClass = "hep-ph",
    doi = "10.1103/PhysRevD.97.023501",
    journal = "Phys. Rev. D",
    volume = "97",
    number = "2",
    pages = "023501",
    year = "2018"
}

@article{Byrnes:2018txb,
    author = "Byrnes, Christian T. and Cole, Philippa S. and Patil, Subodh P.",
    title = "{Steepest growth of the power spectrum and primordial black holes}",
    eprint = "1811.11158",
    archivePrefix = "arXiv",
    primaryClass = "astro-ph.CO",
    doi = "10.1088/1475-7516/2019/06/028",
    journal = "JCAP",
    volume = "06",
    pages = "028",
    year = "2019"
}

@article{Planck:2018jri,
    author = "Akrami, Y. and others",
    collaboration = "Planck",
    title = "{Planck 2018 results. X. Constraints on inflation}",
    eprint = "1807.06211",
    archivePrefix = "arXiv",
    primaryClass = "astro-ph.CO",
    doi = "10.1051/0004-6361/201833887",
    journal = "Astron. Astrophys.",
    volume = "641",
    pages = "A10",
    year = "2020"
}

@article{Bird:2010mp,
    author = "Bird, Simeon and Peiris, Hiranya V. and Viel, Matteo and Verde, Licia",
    title = "{Minimally Parametric Power Spectrum Reconstruction from the Lyman-alpha Forest}",
    eprint = "1010.1519",
    archivePrefix = "arXiv",
    primaryClass = "astro-ph.CO",
    doi = "10.1111/j.1365-2966.2011.18245.x",
    journal = "Mon. Not. Roy. Astron. Soc.",
    volume = "413",
    pages = "1717--1728",
    year = "2011"
}

@article{Fixsen:1996nj,
    author = "Fixsen, D. J. and Cheng, E. S. and Gales, J. M. and Mather, John C. and Shafer, R. A. and Wright, E. L.",
    title = "{The Cosmic Microwave Background spectrum from the full COBE FIRAS data set}",
    eprint = "astro-ph/9605054",
    archivePrefix = "arXiv",
    doi = "10.1086/178173",
    journal = "Astrophys. J.",
    volume = "473",
    pages = "576",
    year = "1996"
}

@article{Chluba:2012we,
    author = "Chluba, Jens and Erickcek, Adrienne L. and Ben-Dayan, Ido",
    title = "{Probing the inflaton: Small-scale power spectrum constraints from measurements of the CMB energy spectrum}",
    eprint = "1203.2681",
    archivePrefix = "arXiv",
    primaryClass = "astro-ph.CO",
    doi = "10.1088/0004-637X/758/2/76",
    journal = "Astrophys. J.",
    volume = "758",
    pages = "76",
    year = "2012"
}

@article{Chung:1999ve,
    author = "Chung, Daniel J. H. and Kolb, Edward W. and Riotto, Antonio and Tkachev, Igor I.",
    title = "{Probing Planckian physics: Resonant production of particles during inflation and features in the primordial power spectrum}",
    eprint = "hep-ph/9910437",
    archivePrefix = "arXiv",
    doi = "10.1103/PhysRevD.62.043508",
    journal = "Phys. Rev. D",
    volume = "62",
    pages = "043508",
    year = "2000"
}

@article{Kofman:2004yc,
    author = "Kofman, Lev and Linde, Andrei D. and Liu, Xiao and Maloney, Alexander and McAllister, Liam and Silverstein, Eva",
    title = "{Beauty is attractive: Moduli trapping at enhanced symmetry points}",
    eprint = "hep-th/0403001",
    archivePrefix = "arXiv",
    reportNumber = "CITA-04-3, SLAC-PUB-10343, SU-ITP-04-05",
    doi = "10.1088/1126-6708/2004/05/030",
    journal = "JHEP",
    volume = "05",
    pages = "030",
    year = "2004"
}

@article{Romano:2008rr,
    author = "Romano, Antonio Enea and Sasaki, Misao",
    title = "{Effects of particle production during inflation}",
    eprint = "0809.5142",
    archivePrefix = "arXiv",
    primaryClass = "gr-qc",
    doi = "10.1103/PhysRevD.78.103522",
    journal = "Phys. Rev. D",
    volume = "78",
    pages = "103522",
    year = "2008"
}

@article{Barnaby:2009mc,
    author = "Barnaby, Neil and Huang, Zhiqi and Kofman, Lev and Pogosyan, Dmitry",
    title = "{Cosmological Fluctuations from Infra-Red Cascading During Inflation}",
    eprint = "0902.0615",
    archivePrefix = "arXiv",
    primaryClass = "hep-th",
    doi = "10.1103/PhysRevD.80.043501",
    journal = "Phys. Rev. D",
    volume = "80",
    pages = "043501",
    year = "2009"
}

@article{Green:2009ds,
    author = "Green, Daniel and Horn, Bart and Senatore, Leonardo and Silverstein, Eva",
    title = "{Trapped Inflation}",
    eprint = "0902.1006",
    archivePrefix = "arXiv",
    primaryClass = "hep-th",
    doi = "10.1103/PhysRevD.80.063533",
    journal = "Phys. Rev. D",
    volume = "80",
    pages = "063533",
    year = "2009"
}

@article{Barnaby:2009dd,
    author = "Barnaby, Neil and Huang, Zhiqi",
    title = "{Particle Production During Inflation: Observational Constraints and Signatures}",
    eprint = "0909.0751",
    archivePrefix = "arXiv",
    primaryClass = "astro-ph.CO",
    doi = "10.1103/PhysRevD.80.126018",
    journal = "Phys. Rev. D",
    volume = "80",
    pages = "126018",
    year = "2009"
}

@article{Barnaby:2010ke,
    author = "Barnaby, Neil",
    title = "{On Features and Nongaussianity from Inflationary Particle Production}",
    eprint = "1006.4615",
    archivePrefix = "arXiv",
    primaryClass = "astro-ph.CO",
    doi = "10.1103/PhysRevD.82.106009",
    journal = "Phys. Rev. D",
    volume = "82",
    pages = "106009",
    year = "2010"
}

@article{Pearce:2017bdc,
    author = "Pearce, Lauren and Peloso, Marco and Sorbo, Lorenzo",
    title = "{Resonant particle production during inflation: a full analytical study}",
    eprint = "1702.07661",
    archivePrefix = "arXiv",
    primaryClass = "astro-ph.CO",
    reportNumber = "ACFI-T17-03, UMN-TH-3619-17, FTPI-MINN-17-03",
    doi = "10.1088/1475-7516/2017/05/054",
    journal = "JCAP",
    volume = "05",
    pages = "054",
    year = "2017"
}

@article{Planck:2018vyg,
    author = "Aghanim, N. and others",
    collaboration = "Planck",
    title = "{Planck 2018 results. VI. Cosmological parameters}",
    eprint = "1807.06209",
    archivePrefix = "arXiv",
    primaryClass = "astro-ph.CO",
    doi = "10.1051/0004-6361/201833910",
    journal = "Astron. Astrophys.",
    volume = "641",
    pages = "A6",
    year = "2020",
    note = "[Erratum: Astron.Astrophys. 652, C4 (2021)]"
}

@article{Blas:2011rf,
    author = "Blas, Diego and Lesgourgues, Julien and Tram, Thomas",
    title = "{The Cosmic Linear Anisotropy Solving System (CLASS) II: Approximation schemes}",
    eprint = "1104.2933",
    archivePrefix = "arXiv",
    primaryClass = "astro-ph.CO",
    reportNumber = "CERN-PH-TH-2011-082, LAPTH-010-11",
    doi = "10.1088/1475-7516/2011/07/034",
    journal = "JCAP",
    volume = "07",
    pages = "034",
    year = "2011"
}

@article{Lesgourgues:2011re,
    author = "Lesgourgues, Julien",
    title = "{The Cosmic Linear Anisotropy Solving System (CLASS) I: Overview}",
    eprint = "1104.2932",
    archivePrefix = "arXiv",
    primaryClass = "astro-ph.IM",
    month = "4",
    year = "2011"
}

@article{Zeldovich:1967lct,
    author = "Zel'dovich, Ya. B. and Novikov, I. D.",
    title = "{The Hypothesis of Cores Retarded during Expansion and the Hot Cosmological Model}",
    journal = "Sov. Astron.",
    volume = "10",
    pages = "602",
    year = "1967"
}

@article{Hawking:1971ei,
    author = "Hawking, Stephen",
    title = "{Gravitationally collapsed objects of very low mass}",
    doi = "10.1093/mnras/152.1.75",
    journal = "Mon. Not. Roy. Astron. Soc.",
    volume = "152",
    pages = "75",
    year = "1971"
}

@article{Carr:1974nx,
    author = "Carr, Bernard J. and Hawking, S. W.",
    title = "{Black holes in the early Universe}",
    doi = "10.1093/mnras/168.2.399",
    journal = "Mon. Not. Roy. Astron. Soc.",
    volume = "168",
    pages = "399--415",
    year = "1974"
}

@article{Carr:1975qj,
    author = "Carr, Bernard J.",
    title = "{The Primordial black hole mass spectrum}",
    doi = "10.1086/153853",
    journal = "Astrophys. J.",
    volume = "201",
    pages = "1--19",
    year = "1975"
}

@article{Sasaki:2018dmp,
    author = "Sasaki, Misao and Suyama, Teruaki and Tanaka, Takahiro and Yokoyama, Shuichiro",
    title = "{Primordial black holes{\textemdash}perspectives in gravitational wave astronomy}",
    eprint = "1801.05235",
    archivePrefix = "arXiv",
    primaryClass = "astro-ph.CO",
    doi = "10.1088/1361-6382/aaa7b4",
    journal = "Class. Quant. Grav.",
    volume = "35",
    number = "6",
    pages = "063001",
    year = "2018"
}

@article{Carr:2020gox,
    author = "Carr, Bernard and K{\"u}hnel, Florian",
    title = "{Primordial Black Holes as Dark Matter: Recent Developments}",
    eprint = "2006.02838",
    archivePrefix = "arXiv",
    primaryClass = "astro-ph.CO",
    doi = "10.1146/annurev-nucl-050520-125911",
    journal = "Ann. Rev. Nucl. Part. Sci.",
    volume = "70",
    pages = "355--394",
    year = "2020"
}

@article{Carr:2020xqk,
    author = "Carr, Bernard and Kohri, Kazunori and Sendouda, Yuuiti and Yokoyama, Jun'ichi",
    title = "{Constraints on primordial black holes}",
    eprint = "2002.12778",
    archivePrefix = "arXiv",
    primaryClass = "astro-ph.CO",
    doi = "10.1088/1361-6633/ac1e31",
    journal = "Rept. Prog. Phys.",
    volume = "84",
    number = "11",
    pages = "116902",
    year = "2021"
}

@article{Green:2020jor,
    author = "Green, Anne M. and Kavanagh, Bradley J.",
    title = "{Primordial Black Holes as a dark matter candidate}",
    eprint = "2007.10722",
    archivePrefix = "arXiv",
    primaryClass = "astro-ph.CO",
    doi = "10.1088/1361-6471/abc534",
    journal = "J. Phys. G",
    volume = "48",
    number = "4",
    pages = "043001",
    year = "2021"
}

@article{Pizzati:2024prz,
    author = "Pizzati, Elia and Hennawi, Joseph F. and Schaye, Joop and Eilers, Anna-Christina and Huang, Jiamu and Schindler, Jan-Torge and Wang, Feige",
    title = "{{\textquoteleft}Little red dots{\textquoteright} cannot reside in the same dark matter haloes as comparably luminous unobscured quasars}",
    eprint = "2409.18208",
    archivePrefix = "arXiv",
    primaryClass = "astro-ph.GA",
    doi = "10.1093/mnras/staf660",
    journal = "Mon. Not. Roy. Astron. Soc.",
    volume = "539",
    number = "4",
    pages = "2910--2925",
    year = "2025"
}

@article{Ma:2025wmv,
    author = "Ma, Yilun and others",
    title = "{No Luminous Little Red Dots: A Sharp Cutoff in Their Luminosity Function}",
    eprint = "2509.02662",
    archivePrefix = "arXiv",
    primaryClass = "astro-ph.GA",
    doi = "10.3847/2041-8213/aea071",
    journal = "Astrophys. J. Lett.",
    volume = "1008",
    number = "2",
    pages = "L64",
    year = "2026"
}

@article{Leach:2001zf,
    author = "Leach, Samuel M. and Sasaki, Misao and Wands, David and Liddle, Andrew R.",
    title = "{Enhancement of superhorizon scale inflationary curvature perturbations}",
    eprint = "astro-ph/0101406",
    archivePrefix = "arXiv",
    doi = "10.1103/PhysRevD.64.023512",
    journal = "Phys. Rev. D",
    volume = "64",
    pages = "023512",
    year = "2001"
}

@article{Kinney:2005vj,
    author = "Kinney, William H.",
    title = "{Horizon crossing and inflation with large eta}",
    eprint = "gr-qc/0503017",
    archivePrefix = "arXiv",
    doi = "10.1103/PhysRevD.72.023515",
    journal = "Phys. Rev. D",
    volume = "72",
    pages = "023515",
    year = "2005"
}

@article{Begelman:2006db,
    author = "Begelman, Mitchell C. and Volonteri, Marta and Rees, Martin J.",
    title = "{Formation of supermassive black holes by direct collapse in pregalactic halos}",
    eprint = "astro-ph/0602363",
    archivePrefix = "arXiv",
    doi = "10.1111/j.1365-2966.2006.10467.x",
    journal = "Mon. Not. Roy. Astron. Soc.",
    volume = "370",
    pages = "289--298",
    year = "2006"
}

@article{Dekker:2024nkb,
    author = "Dekker, Ariane and Kravtsov, Andrey",
    title = "{Constraints on blue and red tilted primordial power spectra using dwarf galaxy properties}",
    eprint = "2407.04198",
    archivePrefix = "arXiv",
    primaryClass = "astro-ph.CO",
    doi = "10.1103/PhysRevD.111.063516",
    journal = "Phys. Rev. D",
    volume = "111",
    number = "6",
    pages = "063516",
    year = "2025"
}

@article{Ando:2024ghr,
    author = "Ando, Shin'ichiro and Balaji, Shyam and Fairbairn, Malcolm and Hiroshima, Nagisa and Ishiwata, Koji",
    title = "{Supermassive black holes from inflation constrained by dark matter substructure}",
    eprint = "2408.11098",
    archivePrefix = "arXiv",
    primaryClass = "astro-ph.CO",
    doi = "10.1103/ybdn-pr3h",
    journal = "Phys. Rev. D",
    volume = "112",
    number = "10",
    pages = "103034",
    year = "2025"
}

@article{Sabti:2021unj,
    author = "Sabti, Nashwan and Mu{\~n}oz, Julian B. and Blas, Diego",
    title = "{New Roads to the Small-scale Universe: Measurements of the Clustering of Matter with the High-redshift UV Galaxy Luminosity Function}",
    eprint = "2110.13161",
    archivePrefix = "arXiv",
    primaryClass = "astro-ph.CO",
    reportNumber = "KCL-2021-75",
    doi = "10.3847/2041-8213/ac5e9c",
    journal = "Astrophys. J. Lett.",
    volume = "928",
    number = "2",
    pages = "L20",
    year = "2022"
}

@article{Munoz:2026diy,
    author = "Mu{\~n}oz, Julian B. and others",
    title = "{A GLIMPSE of Dark Matter: JWST Constraints on Early Structure Formation}",
    eprint = "2610.37751",
    archivePrefix = "arXiv",
    primaryClass = "astro-ph.CO",
    month = "10",
    year = "2026"
}

@article{Inayoshi:2019fun,
    author = "Inayoshi, Kohei and Visbal, Eli and Haiman, Zolt{\'a}n",
    title = "{The Assembly of the First Massive Black Holes}",
    eprint = "1911.05791",
    archivePrefix = "arXiv",
    primaryClass = "astro-ph.GA",
    doi = "10.1146/annurev-astro-120419-014455",
    journal = "Ann. Rev. Astron. Astrophys.",
    volume = "58",
    pages = "27--97",
    year = "2020"
}

@article{Hertzberg:2017dkh,
    author = "Hertzberg, Mark P. and Yamada, Masaki",
    title = "{Primordial Black Holes from Polynomial Potentials in Single Field Inflation}",
    eprint = "1712.09750",
    archivePrefix = "arXiv",
    primaryClass = "astro-ph.CO",
    doi = "10.1103/PhysRevD.97.083509",
    journal = "Phys. Rev. D",
    volume = "97",
    number = "8",
    pages = "083509",
    year = "2018"
}
\end{document}